\documentclass{appolb}
\usepackage{multirow}
\usepackage{graphicx}
\usepackage{epstopdf}
\graphicspath{{./}}
\usepackage{rotating}
\usepackage{geometry}
\usepackage{cite}
\usepackage{amssymb}
\usepackage{pifont}
\usepackage[utf8]{inputenc}
\newcommand{\cmark}{\ding{51}}
\pdfmapline{-dummy FreeSansOblique}
\pdfmapline{-dummy FreeSans}
\pdfmapline{-dummy STIXGeneral}
\begin{document}
\title{
\begin{flushright}
{\bf IFJPAN-IV-2017-12}
\end{flushright}
\vspace*{0.5cm}
Deep learning approach to the Higgs boson CP measurement in $H\to\tau\tau$ decay and associated systematics
}
\author{Elisabetta Barberio$^a$, Brian Le$^{a,b}$, Elzbieta Richter-Was$^c$, Zbigniew Was$^{b,d}$, Daniele Zanzi$^a$, Jakub Zaremba$^{b}$
\address{$^a$ School of Physics, The University of Melbourne, Parkville 3010, Victoria, Australia\\
$^b$ Institute of Nuclear Physics, PAN, Krakow, ul. Radzikowskiego 152, Poland\\
$^c$ Institute of Physics, Jagellonian University, \L{}ojasiewicza 11, 30-348 Krakow, Poland\\
$^d$ CERN, PH-TH, CH-1211 Geneva 23, Switzerland}
}
\maketitle
\begin{abstract}
The $H\to\tau\tau$ decays form the prime channel for the measurement of the Higgs boson state and tests of the CP invariance of Higgs boson couplings. 
A previous study has shown the viability of deep learning techniques for the measurement. 
In this paper, the study is expanded.
Effects due to the partial modelling of experimental effects are discussed.
Furthermore, systematics due to $\tau$ decay modelling for complex cascade decays to $\tau^{\pm}\to a_1^{\pm}\nu_{\tau}\to\rho^{0}\pi^{\pm}\nu_\tau\to3\pi^{\pm}\nu_{\tau}$ are also addressed. 
Various parameterisations are considered using low-energy collision data.
\end{abstract}
\vfill
{\small
\begin{flushleft}
{\bf {IFJPAN-IV-2017-12\\ June 2017}}
\end{flushleft}
}
\newpage
\section{Introduction}
The measurement of the Higgs boson CP state is a fundamental result in the process of establishing the nature of the Higgs boson. 
Possibilities of different spin-CP hypotheses have previously been explored by ATLAS and CMS experiments at the LHC in decays to pairs of vector bosons. 
These measurements have excluded various spin 2 and spin-1 hypotheses and have put an exclusion limit on the spin-CP of $0^-$ (pure pseudoscalar) hypothesis \cite{Aad:2013xqa, Khachatryan:2014kca}.\\ \\
A pseudoscalar (CP odd) Higgs boson is predicted by several theoretical models of new physics such as Supersymmetric and Two Higgs Doublet models \cite{Djouadi:2005gi, Djouadi:2005gj, Accomando:2006ga}.
While the pure pseudoscalar Higgs boson is currently strongly disfavoured, a pseudoscalar Higgs boson could potentially be degenerate with the scalar (CP even) Higgs boson of the Standard Model. 
This scenario would produce a mixed CP state.
Importantly, these measurements mentioned above, which set exclusion limits on the pure states, make use of only the bosonic decay modes of the Higgs ($H\to\gamma\gamma$, $H\to ZZ^{*}$ and $H\to WW^{*}$). 
However, these decays are not sensitive, at tree level, to the possibility of a mixed CP state Higgs boson. 
This can be measured at tree level in decays to fermions, which can couple democratically to both a scalar and a pseudoscalar Higgs boson.\\ \\
The prime candidate for a measurement of the CP mixing of the Higgs boson in decays to fermions, is via $H\to\tau\tau$ decays. 
Several proposals for the measurement exist in the literature \cite{Kramer:1993jn,Bower:2002zx,Desch:2003rw,Rouge:2005iy,Berge:2015nua}. 
The use of deep learning to enhance the sensitivity by including high multiplicity hadronic tau decays was shown to be effective in \cite{Jozefowicz:2016kvz}. 
This paper expands the study outlined in \cite{Jozefowicz:2016kvz} and explores the effect of possible systematic uncertainties (associated with the modelling of $\tau$ decays) and potential degradation of sensitivity due to experimental effects. 
Only $H\to\tau\tau$ decays, in which at least one $\tau$ decays via $\tau\to a_1\nu$, will be considered.
\section{CP Sensitive Observables}
The construction of a simple CP sensitive observable is well established in literature \cite{Kramer:1993jn,Bower:2002zx,Desch:2003rw,Rouge:2005iy,Berge:2015nua}. 
A mixture of scalar (CP even - $\phi_\tau=0$) and pseudoscalar (CP odd - $\phi_\tau=\pi/2$) Higgs boson couplings to $\tau$ leptons can be expressed in the Lagrangian
\begin{equation}
\mathcal{L}_{int} = g_{\tau}\overline{\tau}(\cos\phi_{\tau} + \sin\phi_{\tau}i\gamma_5)\tau h,
\end{equation}
where $\phi_{\tau}$ is the mixing angle that parameterises the relative strengths of the scalar and pseudoscalar couplings.
The mixing angle is exposed in the Higgs boson decay width via the transverse spin components of the $\tau$ leptons 
\begin{equation}
\Gamma (h_{mix}\to\tau^{+}\tau^{-}) \sim 1 - s^{\tau^{+}}_{\parallel}s^{\tau^{-}}_{\parallel} + s^{\tau^{+}}_{\bot}R(2\phi_{\tau})s^{\tau^{-}}_{\bot},
\end{equation}
where $R$ is a rotation in the x-y (transverse) plane \cite{Desch:2003rw} and $s^{\tau^{+}}_{\parallel}$ and $s^{\tau^{+}}_{\bot}$ are the components of the $\tau$ spin in the direction parallel and perpendicular to the $\tau$ momenta, in the rest frame of the Higgs, respectively.\\ \\
Ultimately, thanks to the maximally parity violating nature of $\tau$ lepton decays, the mixing angle is observable through the angular distributions of the $\tau$ lepton decay products.\\ \\
In the simple case of $\tau$ lepton decays via a $\rho^{\pm}$ resonance, the acoplanarity angle is defined as the angle between the planes spanned from the decay products (one charged and one neutral pion) of both $\tau$ leptons. 
This has been shown to be a observable which is sensitive to the CP state of the Higgs boson\cite{Bower:2002zx,Desch:2003rw}. 
For more complex decay modes, such as $\tau^{\pm}\to a_1^{\pm} \nu_\tau$, this becomes further complicated due to the increasing number of relevant variables which can be formed in the cascade decay ($a_1^{\pm} \nu_\tau\to\rho^{0}\pi^{\pm}\nu_\tau\to 3\pi\nu_\tau$). 
This creates a highly multidimensional problem which is challenging to account for correlations between relevant variables. 
In \cite{Jozefowicz:2016kvz}, a neural network approach was shown to be effective in encompassing the CP sensitive observable information in a classifier score.
\section{Neural Network Approach}
In the following, a similar neural network (NN) setup to the one developed in \cite{Jozefowicz:2016kvz} will be adopted. The Monte-Carlo simulations (MC) used as a baseline is the same as generated for results derived in\cite{Jozefowicz:2016kvz}. A brief description of the samples, inputs and NN setup is summarised below.\\ \\
The $H\to\tau\tau$ events, produced via gluon-gluon fusion, were simulated with {\tt Pythia 8.2}.
A total of 2.5 million and 5 million events are generated for $H\to\tau\tau\to a_1^{\pm}\nu\ a_1^{\mp}\nu$ ($a_1-a_1$) and $H\to\tau\tau\to\rho^{\pm}\nu\ a_1^{\mp}\nu$ ($a_1-\rho$) decays respectively.
From the events generated, training samples consisting of an initial sample of 1 million events (prior to filtering) and the validation and test samples containing approximately 300,000 and 400,000 events (after filtering) were used for the $a_1-\rho$ and $a_1-a_1$ modes respectively
\footnote{Due to filtering of events based on kinematic selections, not all events are ultimately used.}.
The spin correlations were implemented with {\tt TauSpinner} \cite{Przedzinski:2014pla} weights and decays were simulated with the {\tt TAUOLA} library \cite{Davidson:2010rw}.\\ \\
For each sample events, sets of input features are calculated as inputs for the NNs. 
Combinations of potentially important features were included:
\begin{itemize}
\item $\phi^{*}$ - The acoplanarity angle, the basic CP sensitive variable \cite{Bower:2002zx}. This is defined as the angle between two planes (which can be formed from either two pions or three pions where two have been summed to a $\rho^{0}$).
\item $y$ - An essential variable used to separate events into two categories which reveals modulations in the acoplanarity angle \cite{Bower:2002zx}. This is defined as $\frac{E_{\pi^{\pm}} - E_{\pi^{0}}}{E_{\pi^{\pm}} + E_{\pi^{0}}}$ for decays of intermediate $\rho^{\pm}$ resonances (a similar definition for $\rho^{0}$ resonances). For decays of $a_1 \to\rho^{0}\pi^{\pm}$ this must be modified (due to the large mass of the $\rho^{0}$ resonance) to be $\frac{E_{\pi^{\pm}} - E_{\pi^{0}}}{E_{\pi^{\pm}} + E_{\pi^{0}}} - \frac{m^{2}_{a_1} - m^{2}_{\pi^{\pm}} + m^{2}_{\rho^{0}}}{2m^{2}_{a_1}}$
\item $m_i$- Invariant masses of pairs of detectable decay products (also triplets in the case of $a_1$ decays). This can be useful in the case of $a_1$ decays due to ambiguities in which pions form intermediate resonances in the cascade decay.
\item 4-vectors - The four-momenta of the outgoing pions calculated in the frame of the decay planes. In principle, this class of features should contain an equivalent set of information as the other features mentioned here.
\end{itemize}
The neural networks consist of six layers of 300 nodes each with a single output (the classifier score), and uses the Adam optimiser \cite{DBLP:journals/corr/KingmaB14}. 
The number of training iterations (epochs) was increased (from 5 in \cite{Jozefowicz:2016kvz}) to 50 and 70 for decays via the $a_1\rho$ and $a_1a_1$ modes respectively. This ensures a well trained neural network is produced. 
Additionally the batch size (between 100 and 500) and dropout (ranging between 10\% and 30\%) was optimised for the decay mode and set of input features. 
The minimal amount of dropout (whilst still maintaining a well-trained non-trivial network) was used in training. The batch normalisation layers, used in \cite{Jozefowicz:2016kvz},  were removed from the networks.\\ \\
The AUC (area under the curve) of the ROC (receiver operator characteristic) curve has been used as a figure of merit for the separation power of neural network \cite{roc}. 
This measure accounts for both the behaviour of the true positives and the false positives in the classifier score.\\ \\

\section{Experimental and Theoretical Considerations}
When discussing possible systematic effects which can be a detriment to the separation, one should consider a mix of experimental and theoretical effects. 
From the experimental side, the detector resolution limits the precision at which the outgoing pion four-momenta can be measured. 
This is addressed in simulation by Gaussian smearing of the momenta components based on detector resolutions representative of the ATLAS detector \cite{Aad:2010bx,Aad:2015unr}. 
As an example of theoretical considerations, we investigate systematic effects from the modelling of the $\tau$ lepton decay itself, a necessity in dealing with complex cascade decays. 
Both effects will be assessed in the following subsections.

\subsection{Detector Resolution Effects}
As a baseline, NN are trained with different combinations of input features with exact MC. 
However, these exact MC are not indicative of the conditions of actual detectors at the LHC experiments. 
Detector resolution limits the precision at which the momenta are measurable. 
This may impact the stability of the NN approach if the NN is too sensitive to detector resolution effects. 
To assess the impact of experimental sources of uncertainty (due to detector resolution effects) smeared samples are produced to mimic the detector response. 
New NNs are trained against these smeared samples.\\ \\
Simple Gaussian smearings of input features through the input four-momenta were implemented in order to estimate the impact on the sensitivity. 
For charged pions, the resolutions of $\sigma(\theta) = 0.88$ mrad, $\sigma(\phi) = 0.147$ mrad and $\sigma(1/p$) = 4.83$\times$10$^{-4}$ GeV$^{-1}$ were used to mimic the inner tracking resolution \cite{Aad:2010bx}. 
For neutral pions, which are reconstructed from calorimeter deposits not associated with a track, the resolutions of  $\sigma(\eta) = 0.0056$ rad, $\sigma(\phi) = 0.012$ rad and $\sigma(E_T) = 0.16 \cdot E_T$ were used \cite{Aad:2015unr}.\\ \\
More complex parameterisations of the experimental effects may be ultimately needed, but to the first order only a small effect on the sensitivity of the NN (see Table \ref{table:baseline}) is evident.\\ \\
In Table \ref{table:baseline} statistical and systematic uncertainties on the AUC score are also given. The statistical uncertainty was estimated through a bootstrap method \cite{doi:10.1137/1.9781611970319}. 
In short, the bootstrap method creates a new sample of size N by sampling from the original MC sample N times, each time returning the event back.
This allows for the creation of a number of (only partially correlated) samples without the need for generating new MC samples. 
The systematic uncertainty was estimated from variations due to the procedure of smearing.
The following procedure was implemented to calculate these uncertainties:
\begin{itemize}
\item Create an ensemble of samples with the same size as the original sample.
    \begin{itemize}
    \item For statistical uncertainties, each sample (either smeared or exact) is created by bootstrapping the original MC sample (the new sample is the same size as the original). 1000 such samples were generated.
    \item For systematic uncertainties, 200 identical samples were smeared (with different random seeds) in the lab frame four-momenta of the outgoing pions.
    \end{itemize}
\item Calculate AUC score of each sample in the ensemble. Each AUC score is calculated by applying a NN trained on either exact or smeared MC.
    \begin{itemize}
    \item For the bootstrap (statistical) uncertainty, NNs trained on either exact or smeared MC are used.
    \item For the systematic uncertainty only NNs trained on smeared MC are used.
    \end{itemize}
\item Calculate the mean and width of the distribution of AUC scores\footnote{In rare cases the smearing procedure produces an AUC score which is very far from the mean of the distribution of AUC scores. These cases are removed from the calculation of the mean and width.}. 
\end{itemize}
\begin{table}
    \centering
    \begin{tabular}{|c|c|c|c|c|c|c|c|}
    \hline
    \multicolumn{4}{|c|}{Features}  & \multicolumn{1}{c|}{\multirow{2}{25 mm}{Exact $\pm$ (stat)}} & \multirow{2}{50 mm}{Smeared $\pm$ (stat) $\pm$ (syst)} & \multicolumn{1}{c|}{\multirow{2}{20 mm}{From \cite{Jozefowicz:2016kvz}}} \\ \cline{1-4}
    $\phi^*$ & 4-vec &  $y_i$ & $m_i$ & \multicolumn{1}{c|}{}    & \multicolumn{1}{c|}{}             &                      \\
    \hline
    \multicolumn{7}{|c|}{$a_1-\rho$ Decays}\\
    \hline
    \cmark & \cmark & \cmark & \cmark & $0.6035 \pm 0.0005$ & $0.5923 \pm 0.0005 \pm 0.0002$ & 0.596 \\
    \cmark & \cmark & \cmark & -      & $0.5965 \pm 0.0005$ & $0.5889 \pm 0.0005 \pm 0.0002$ & - \\
    \cmark & \cmark & -      & \cmark & $0.6037 \pm 0.0005$ & $0.5933 \pm 0.0005 \pm 0.0003$ & - \\
    -      & \cmark & -      & -      & $0.5971 \pm 0.0005$ & $0.5892 \pm 0.0005 \pm 0.0002$ & 0.590 \\
    \cmark & \cmark & -      & -      & $0.5971 \pm 0.0005$ & $0.5893 \pm 0.0005 \pm 0.0002$ & 0.594 \\
    \cmark & -      & \cmark & \cmark & $0.5927 \pm 0.0005$ & $0.5847 \pm 0.0005 \pm 0.0002$ & 0.578 \\
    \cmark & -      & \cmark & -      & $0.5819 \pm 0.0005$ & $0.5746 \pm 0.0005 \pm 0.0002$ & 0.569 \\
    \hline
    \multicolumn{7}{|c|}{$a_1-a_1$ Decays}\\
    \hline
    \cmark & \cmark & \cmark & \cmark & $0.5669 \pm 0.0004$ & $0.5657 \pm 0.0004 \pm 0.0001$ & 0.573 \\
    \cmark & \cmark & \cmark & -      & $0.5596 \pm 0.0004$ & $0.5599 \pm 0.0004 \pm 0.0001$ & - \\
    \cmark & \cmark & -      & \cmark & $0.5677 \pm 0.0004$ & $0.5661 \pm 0.0004 \pm 0.0001$ & - \\
    -      & \cmark & -      & -      & $0.5654 \pm 0.0004$ & $0.5641 \pm 0.0004 \pm 0.0001$ & 0.553 \\
    \cmark & \cmark & -      & -      & $0.5623 \pm 0.0004$ & $0.5615 \pm 0.0004 \pm 0.0001$ & 0.573 \\
    \cmark & -      & \cmark & \cmark & $0.5469 \pm 0.0004$ & $0.5466 \pm 0.0004 \pm 0.0001$ & 0.548 \\
    \cmark & -      & \cmark & -      & $0.5369 \pm 0.0004$ & $0.5374 \pm 0.0004 \pm 0.0001$ & 0.536 \\
    \hline
    \end{tabular}
    \caption{AUC for NN trained to separate scalar and pseudoscalar hypotheses with combinations of input features marked with a \cmark. 
    Results in the column labelled ``Exact" are from NNs trained with exact sample. 
    The results in column labelled ``Smeared" are from NNs trained with smeared sample. 
    Statistical uncertainties are derived from a bootstrap method described in the text. 
    Systematic uncertainty is calculated with the method described in the text.}
    \label{table:baseline}
\end{table}
Results are summarised in Table \ref{table:baseline} and will be used as a reference for considerations of further systematic effects. 
The degradation in sensitivity due to training on smeared samples is of 1\% level significance. 
Additionally tests applying NN (trained on smeared samples) to exact samples demonstrated a smaller difference than present between the columns in Table \ref{table:baseline}.
This demonstrates some robustness of the NN against the smearing procedure. 
The size of the systematic uncertainty due to the smearing procedure is less than 0.04\%, which is dwarfed by the statistical uncertainty (of the order of 0.07-0.08\%).

\subsection{Systematics from $\tau$ Decay Modelling}
Fundamental to the simulation of $\tau$ decays to hadronic final states is the parameterisation of the decay model. 
As these decays lie in a regime of medium energy QCD interactions, they rely often on data-driven parameterisations through various form-factors. 
The modelling is particularly important to the cascade decays $a_1$, which involve the propagation of spin into broad intermediate vector resonances (which are difficult to model).
The parameterisations are based on measurements from low energy experiments such as CLEO \cite{Asner:1999kj} and BaBar \cite{Chrzaszcz:2016fte}.\\ \\
As the polarisation of the $\tau$ may potentially be sensitive to the modelling through these parameterisations, different models are tested in order to assess the impact on the final measurement of the Higgs boson CP state. 
Variations of currents available through the {\tt TAUOLA} package \cite{Davidson:2010rw} 
for the $\tau^- \to \pi^- \pi^- \pi^+ \nu_\tau$ decay were motivated by 
    very specific considerations, both theoretical and experimental.
    Here, we present briefly some comments on their origins:

    \begin{enumerate}
    \item {\bf Standard CLEO (STD)} - a
    parameterisation using the Kuhn-Santamaria (KS) model~\cite{Kuhn:1990ad}, 
    extended by CLEO with resonances in $\pi^0 \pi^-$ spectrum.
    It was developed by CLEO Collaboration for the 
    $\tau \to \pi^0 \pi^0 \pi^- \nu_\tau$ decays~\cite{Asner:1999kj}.
    Even though some of postulated resonances were not well established,
    the CLEO collaboration only fit the coefficients while masses and 
    widths were taken from PDG tables and theoretical predictions of that time.
    The strength of this model is the fit to three-dimensional data~\cite{Asner:1999kj}. 
    This parameterisation uses the same 
    modelling for both three- and one-prong (the number of charged pions) channels.
    The projection operators~\cite{Kuhn:1992nz}, which in principle enable
    access to all details of differential distributions, were used for some crosschecks.

    \item {\bf Alternative CLEO (ALT)} 
    - is a variant of the standard parameterisation (STD) obtained by isospin rotation of 
    contributions to current, from the
    $\pi^0 \pi^0 \pi^-$ to the $\pi^- \pi^- \pi^+ $ channel.
    Numerical constants remained unmodified. 
    While it seems this should be superior to the STD model, 
    it lacks experimental publications validating it. 
    This parameterisation is described with additional extensions in~\cite{Hinson},
    but even though it is more than 15 years old, it 
    never made it to the collaboration publication.
    For the purpose of this study, consideration of hadronic currents as a systematic uncertainty, it is 
    nonetheless sufficiently well established.

    \item {\bf BaBar (BBR)} - a parameterisation used in the BaBar Collaboration based on the 
        KS model~\cite{Kuhn:1990ad} without any extensions developed by CLEO. 
        This parameterisation relies on measurements from a much larger dataset than CLEO.
        However, it must be noted this dataset is obtained from collisions 
        in which the $\tau$ leptons must be considered relativistic. 
        The invariant masses of $\pi^-\pi^-\pi^+$ and $\pi^-\pi^+$ 
        systems~\cite{Nugent:2009zz} were used for fits.
        The parameterisation is determined through distributions obtained from
        the collaboration production files, see \cite{Chrzaszcz:2016fte} for details.

        \item {\bf Resonance Chiral Lagrangian (R$\chi$L)} - this parameterisation is motivated by 
        considerations of~\cite{Dumm:2009va}. 
        It is based on fits to invariant mass distributions of $\pi^-\pi^-\pi^+$ and $\pi^-\pi^+$ 
        systems of BaBar measurements~\cite{Nugent:2009zz}. 
        It achieved its present form in ~\cite{Nugent:2013hxa},
        where also $\pi^-\pi^-$ invariant mass information was used for fits.
        This model is inherently different from the KS model; 
        this is of value for discussion of systematic errors.
        A weak point of the model is the lack of comparison with three-dimensional data,
        meaning poor experimental verification.

        \end{enumerate}

\begin{figure}[htb]
    \begin{center}
    {
        \includegraphics[width=6.8cm]{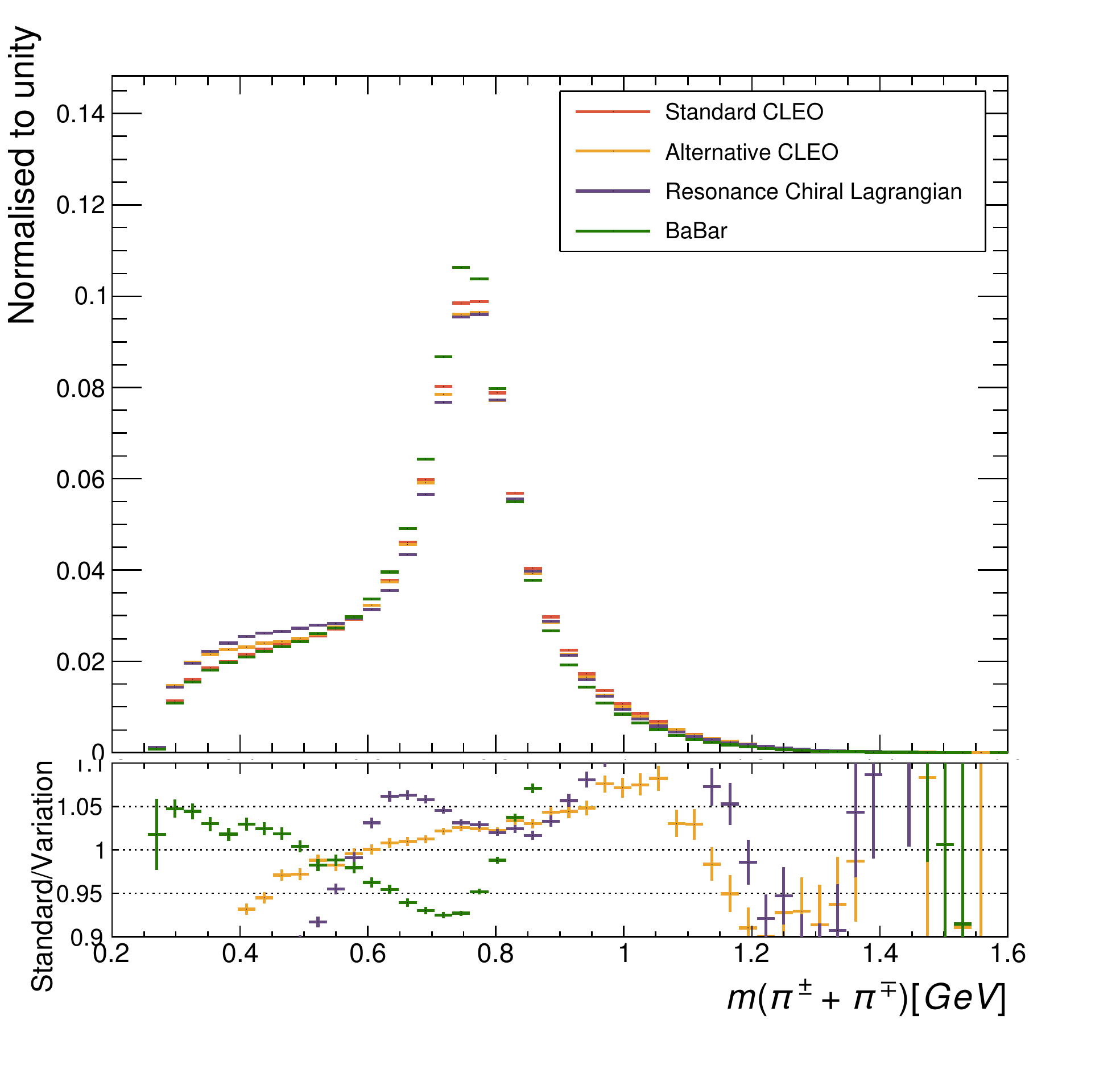}
        \includegraphics[width=6.8cm]{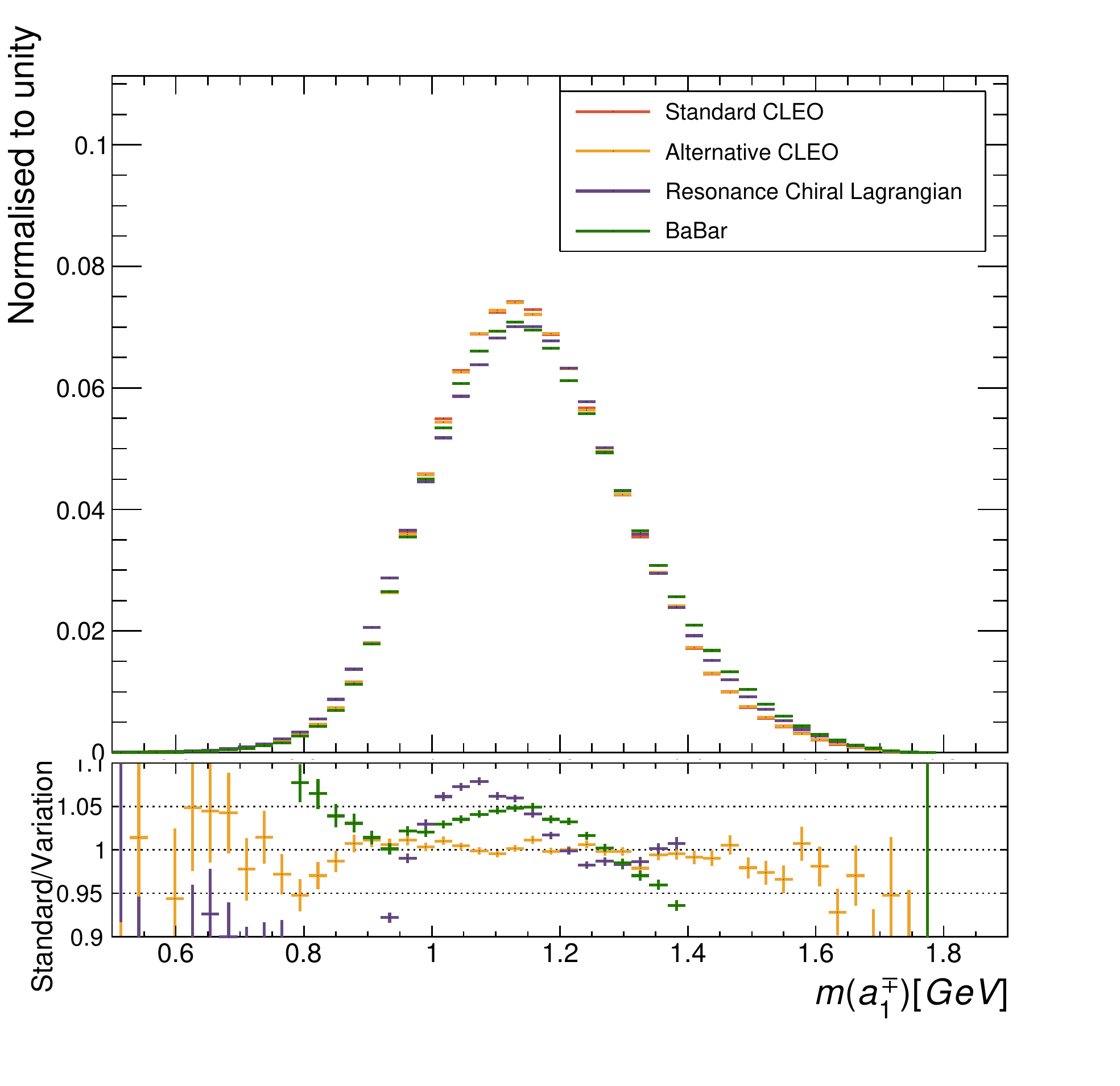}
    }
    \end{center}
    \caption{Invariant masses constructed from $\tau\to a_1\nu\to 3\pi\nu$ decays. 
The lower panels represent the ratios between the alternative current (R$\chi$L, ALT, BBR) and the 
standard (STD) current. Left, the two pion mass   formed from the oppositely charged pions.
Right, mass of  all  pions of the $a_1$ decay  combined.
}
    \label{Fig:mass}
\end{figure}

\begin{figure}[htb]
    \begin{center}
    {
        \includegraphics[width=6.8cm]{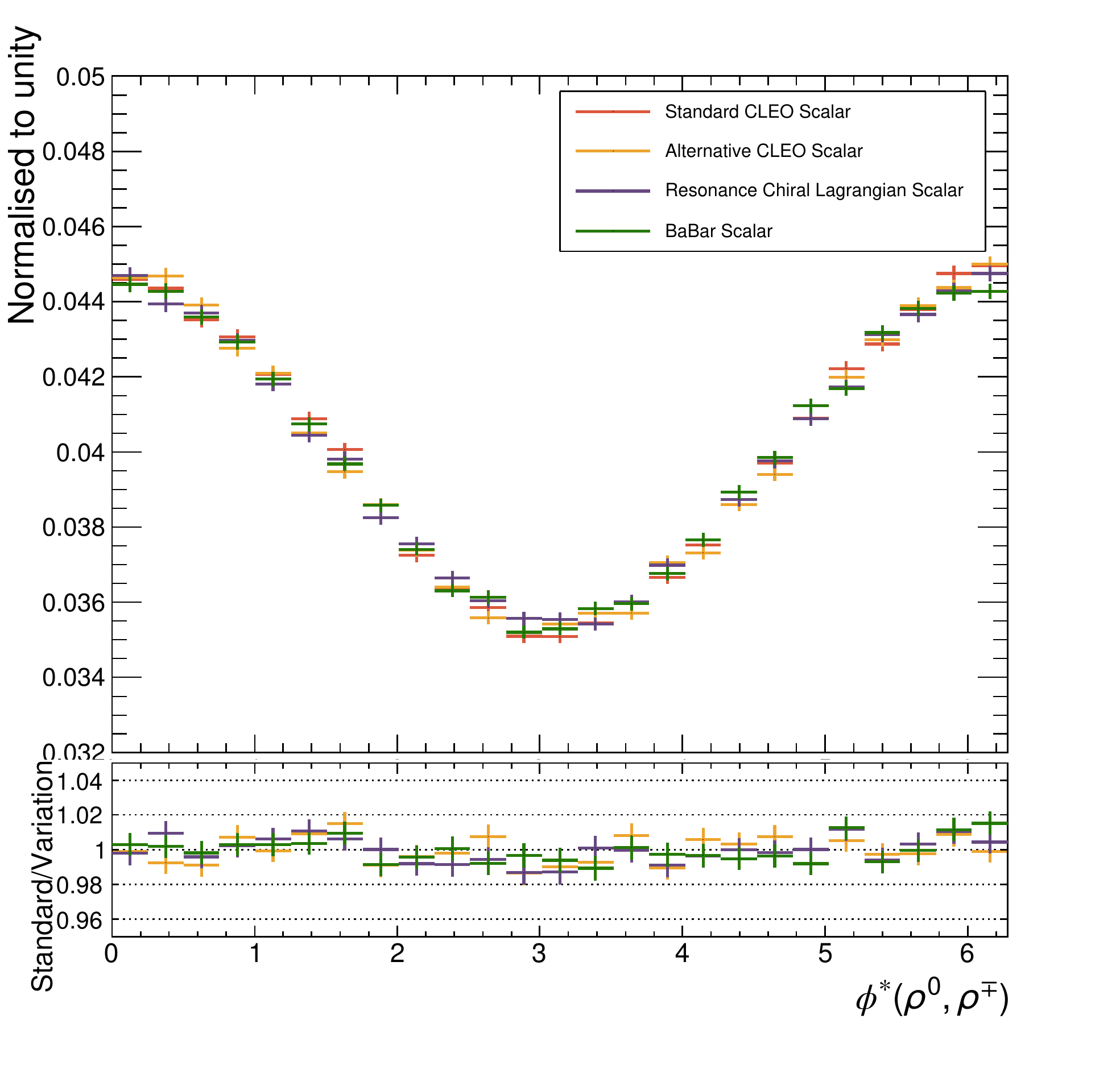}
        \includegraphics[width=6.8cm]{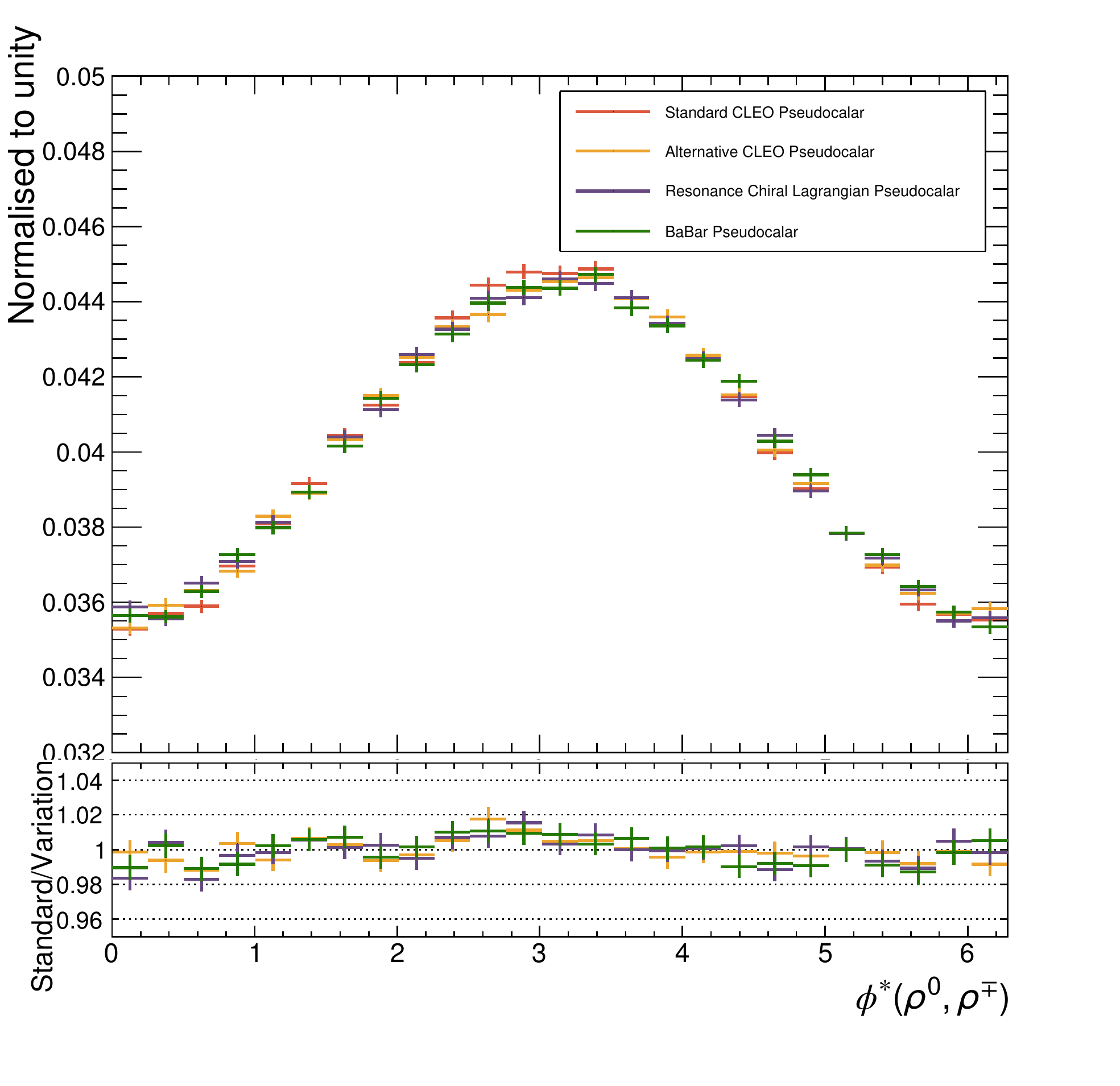}
    }
    {
        \includegraphics[width=6.8cm]{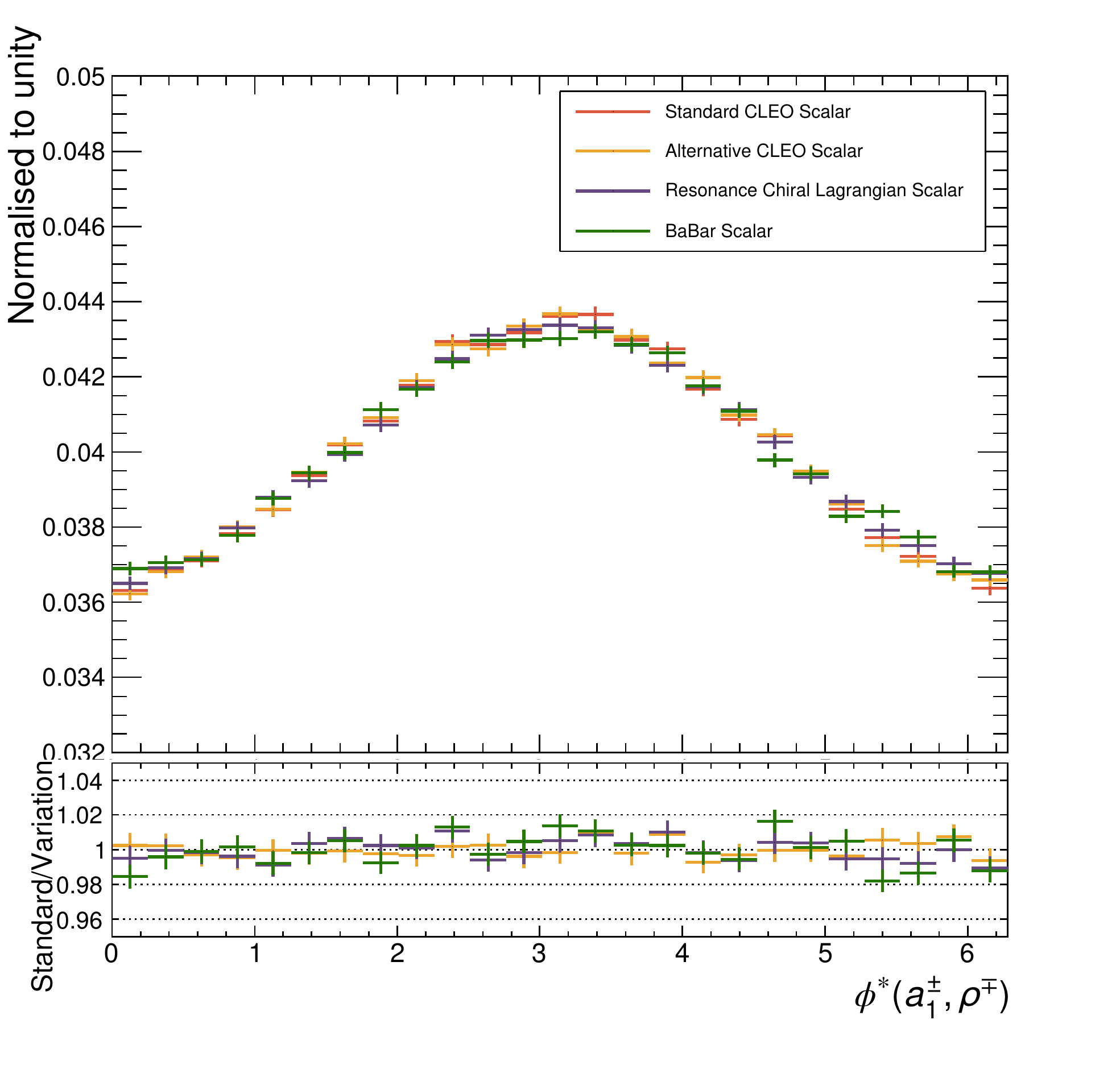}
        \includegraphics[width=6.8cm]{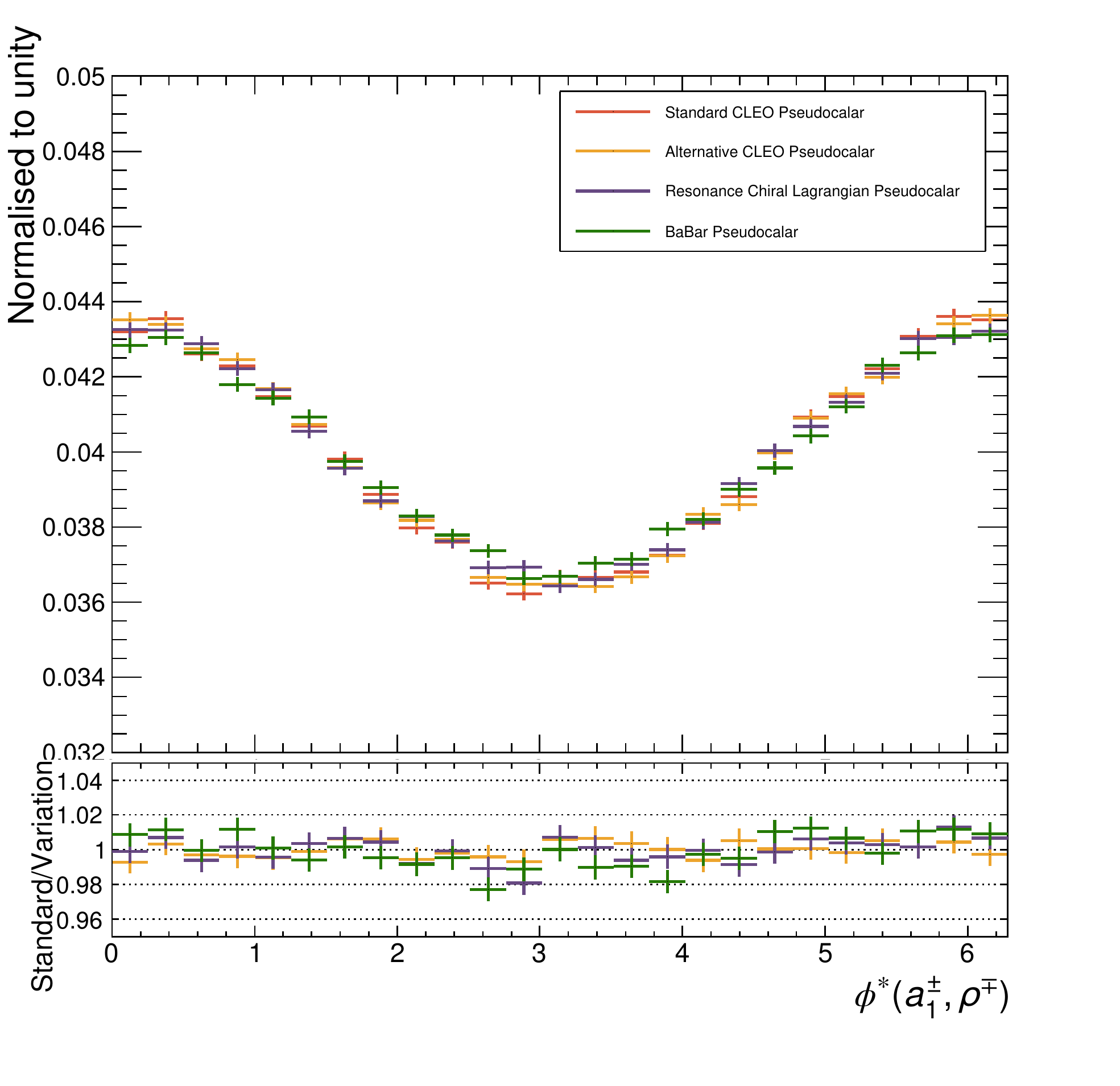}
    }
    \end{center}
    \label{Fig:aco_a1rho_a1rho_pseudo}
    \caption{A comparison of acoplanarity angles calculated for $H\to\tau\tau\to \rho^{\pm}\nu a_1^{\mp}\nu$ decays with different 
parameterisations of the hadronic currents (STD, R$\chi$L, ALT, BBR). The lower panels show the ratios between the alternative currents 
(R$\chi$L, ALT, BBR) and the standard (STD) current. The top and bottom rows contain acoplanarities constructed from combinations 
of reconstructed planes of $\rho^{0}$-$\rho^{\pm}$, $a_1^{\pm}$-$\rho^{\pm}$ respectively. Each row contains the distributions of 
the acoplanarity angle for scalar (left) and pseudoscalar (right) hypotheses for events passing a selection of $y_1\cdot y_2 > 0$.}
    \label{Fig:acoplanar_a1rho}
\end{figure}

\begin{figure}[htb]
    \begin{center}
    {
        \includegraphics[width=6.8cm]{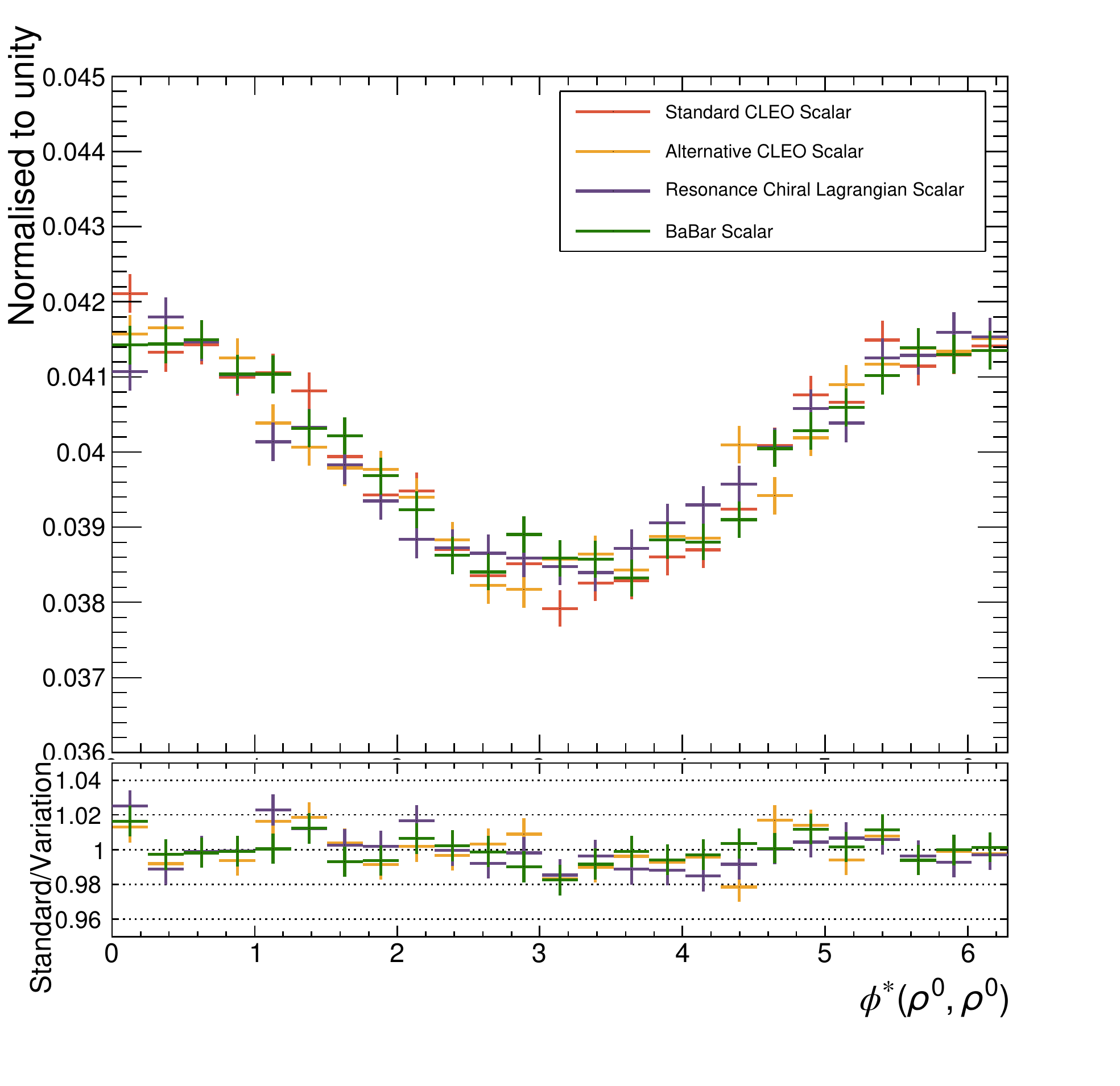}
        \includegraphics[width=6.8cm]{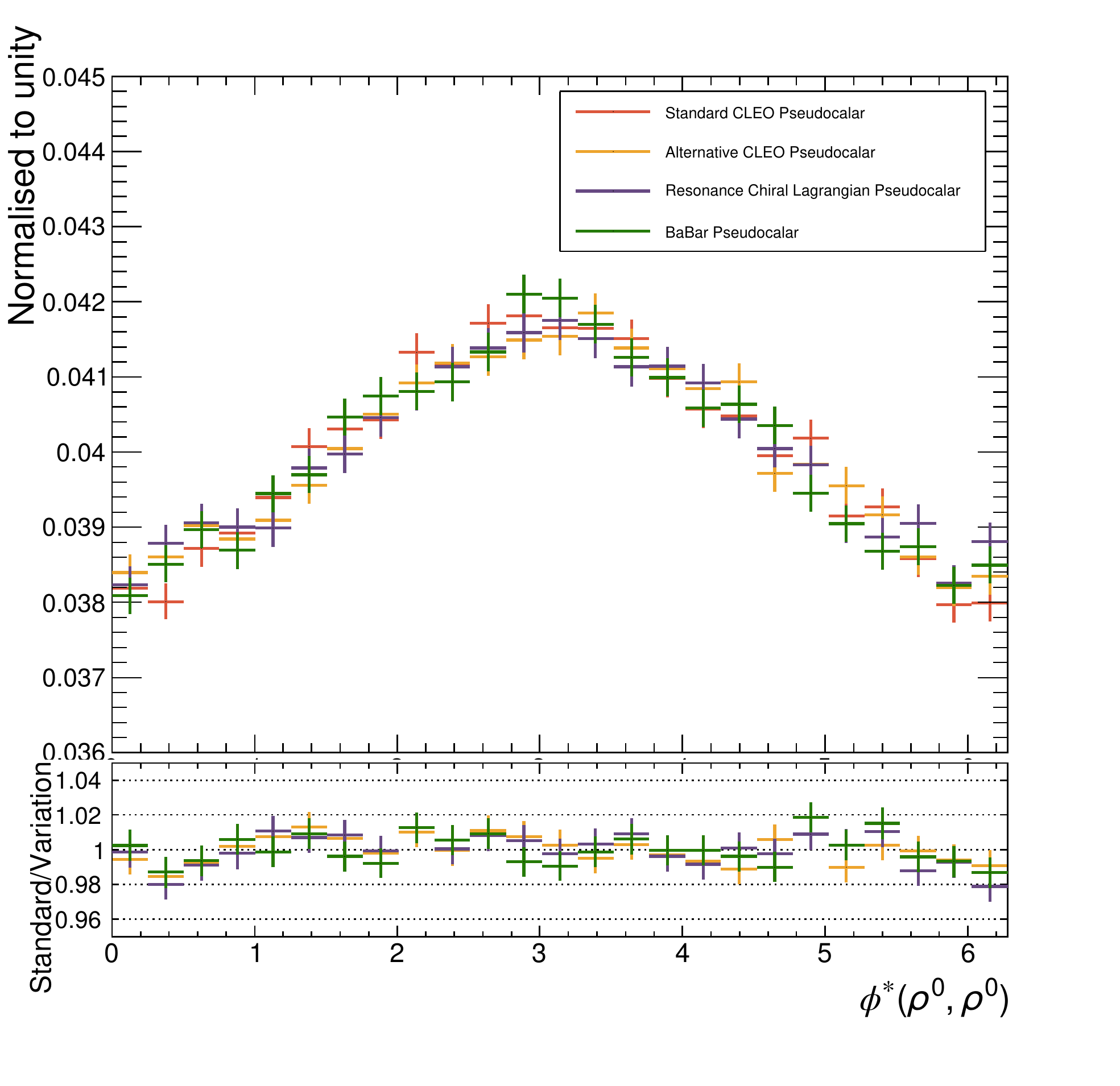}
    }
    {
        \includegraphics[width=6.8cm]{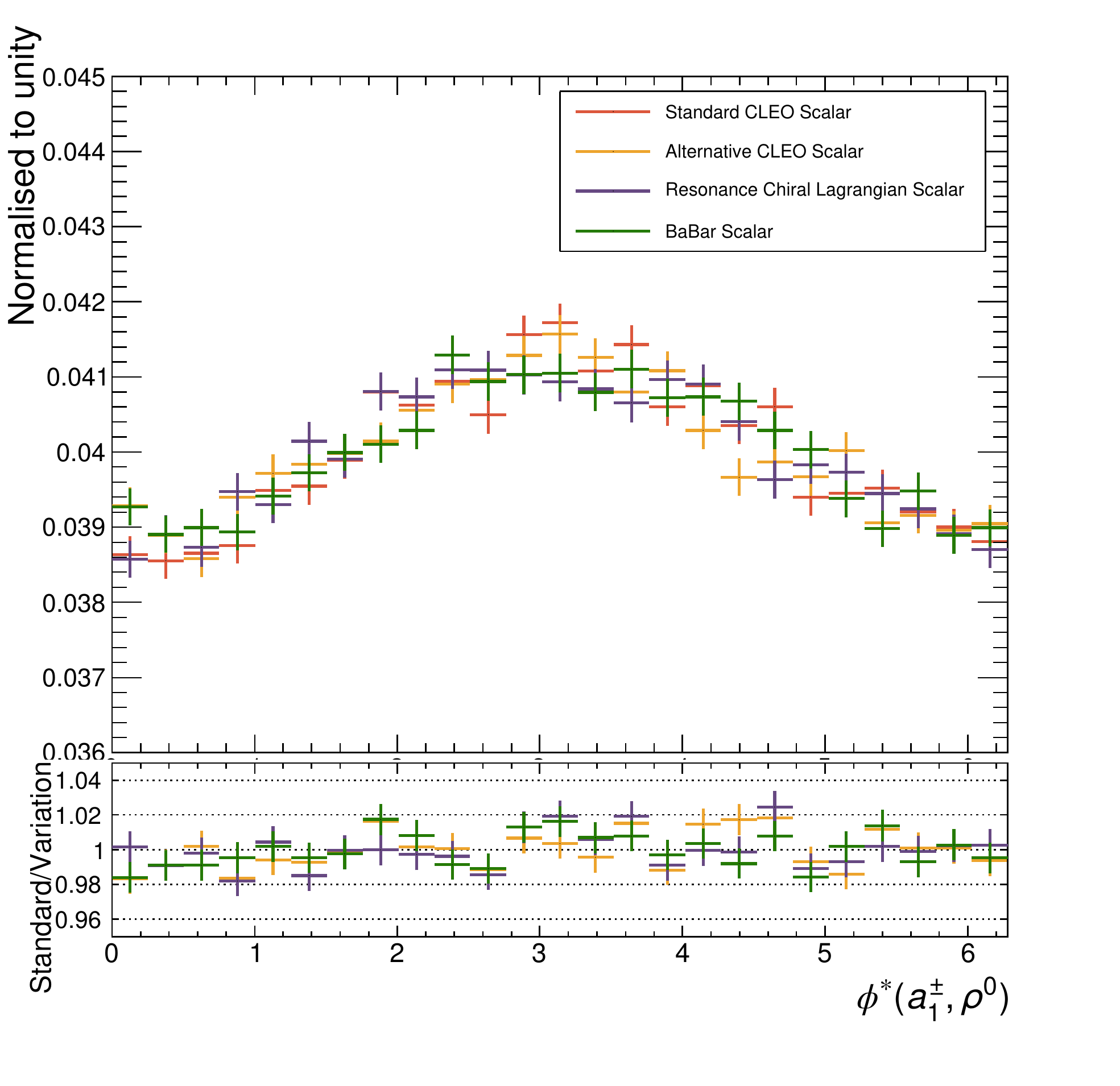}
        \includegraphics[width=6.8cm]{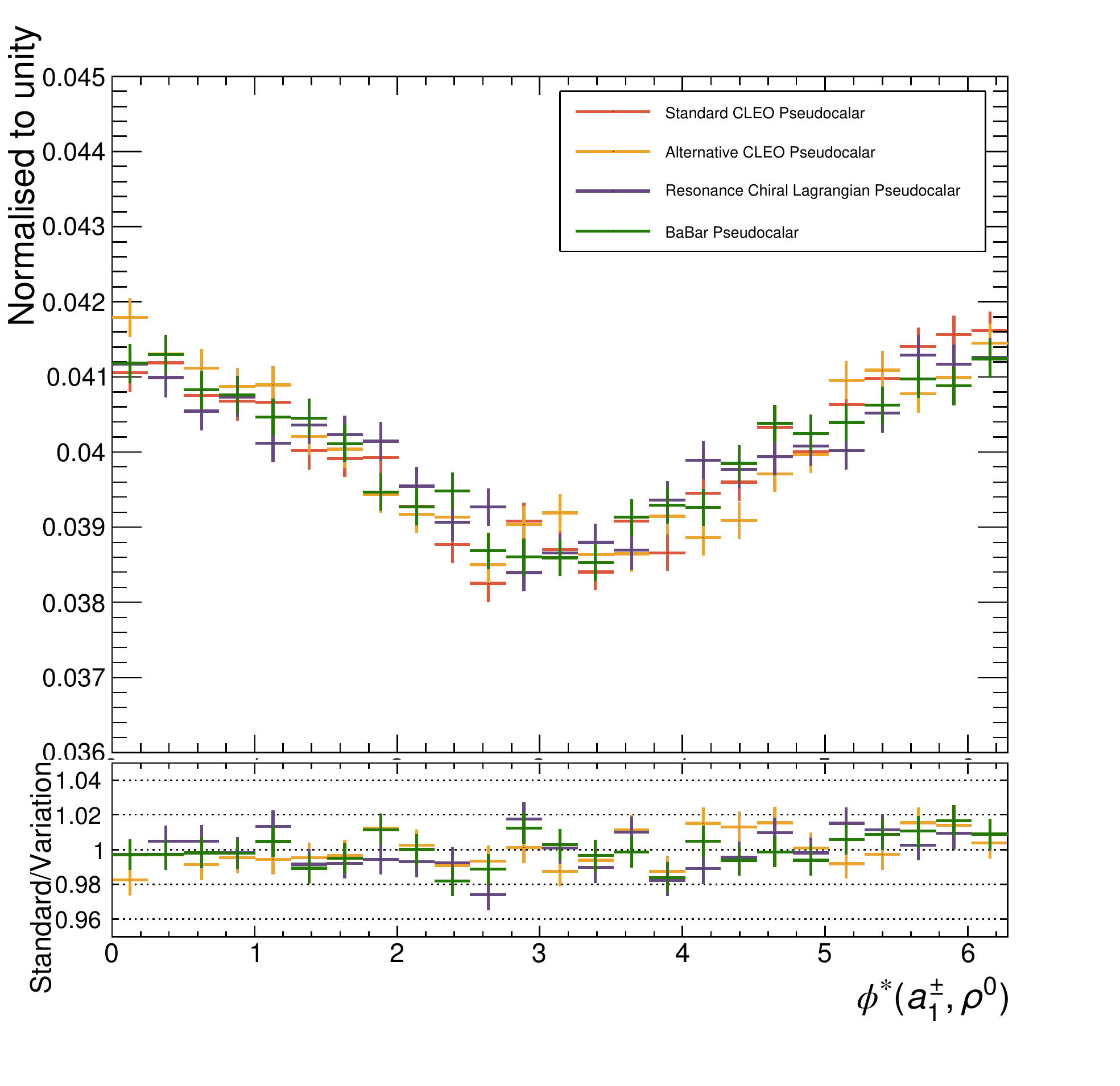}
    }
    {
        \includegraphics[width=6.8cm]{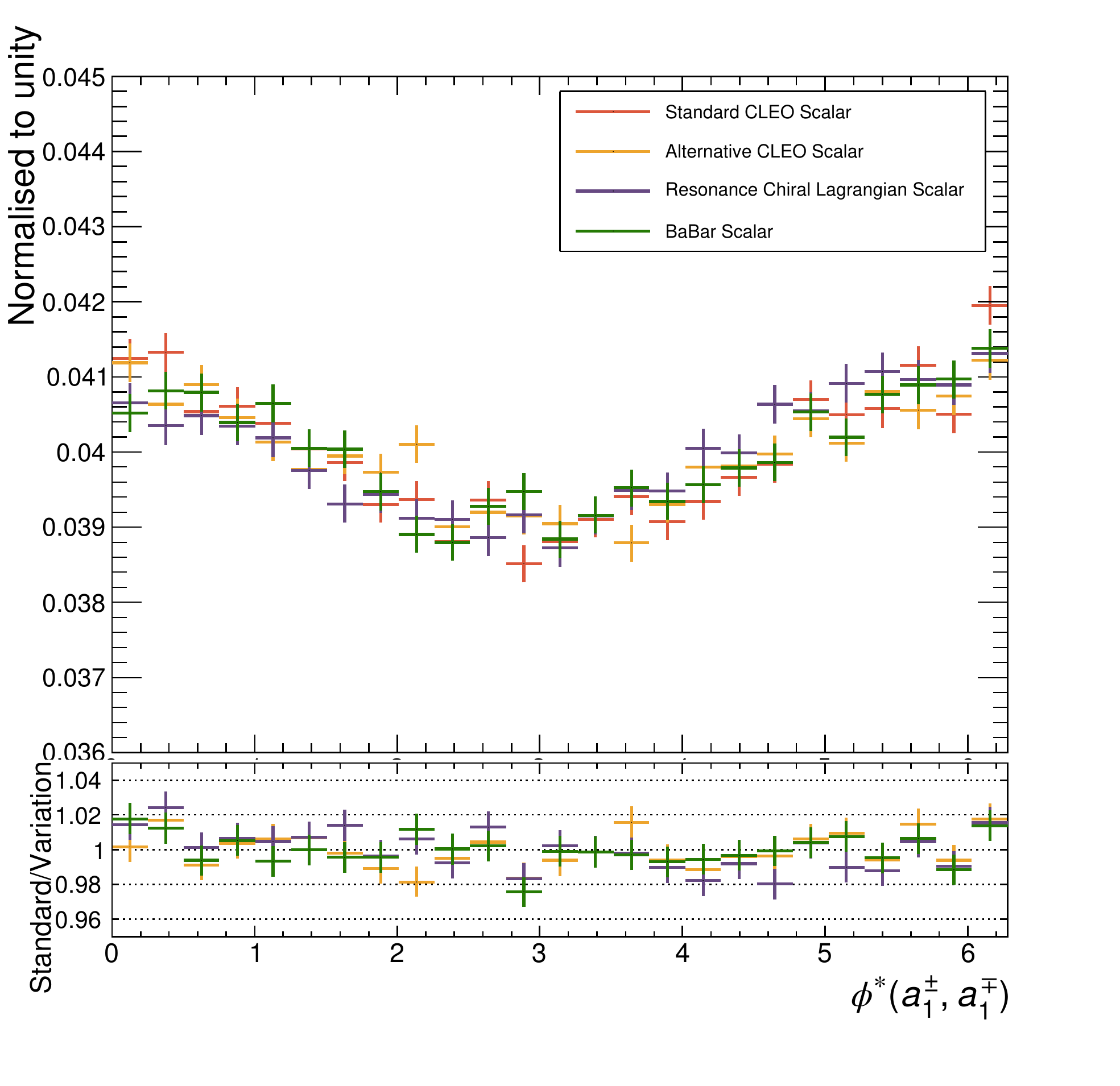}
        \includegraphics[width=6.8cm]{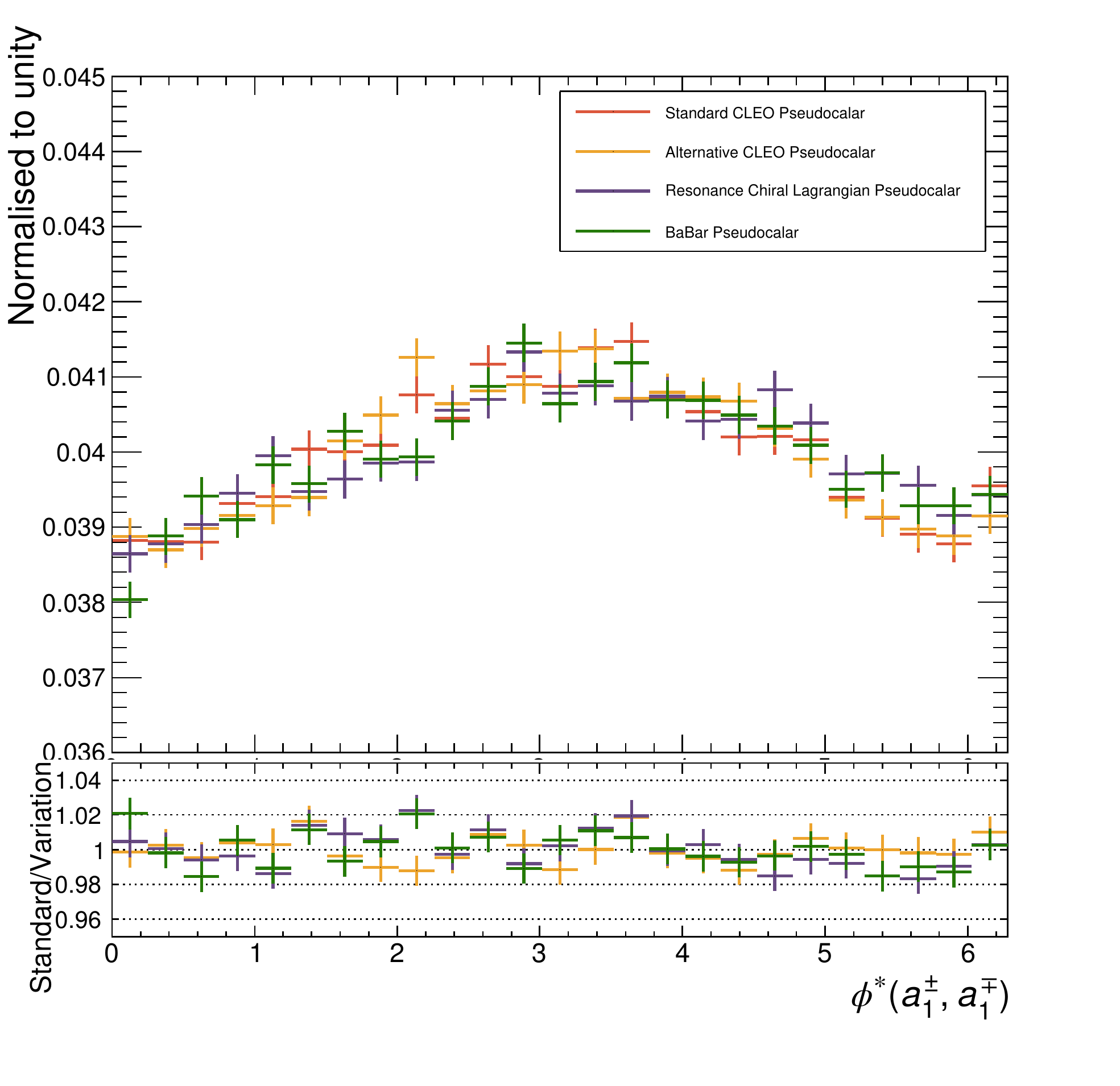}
    }
    \end{center}
    \caption{A comparison of acoplanarity angles calculated for $H\to\tau\tau\to a_1\nu a_1\nu$ decays with different parameterisations 
of the hadronic currents (STD, R$\chi$L, ALT, BBR). The lower panels show the ratios between the alternative currents (R$\chi$L, ALT, BBR) 
and the standard (STD) current. The top, middle and bottom rows contain acoplanarities constructed from combinations of reconstructed 
candidates of $\rho^{0}$-$\rho^{0}$, $a_1^{\pm}$-$\rho^{0}$, and $a_1^{\pm}-a_1^{\pm}$ respectively. Each row contains 
the distributions of the acoplanarity angle for scalar (left) and pseudoscalar (right) hypotheses for events passing a selection of 
$y_1\cdot y_2 > 0$.}
    \label{Fig:acoplanar_a1a1}
\end{figure}

        These are the currents which are presently
        available for tests of systematics in CP measurements with 
        $\tau^{\pm}\to3\pi^{\pm}\nu_\tau$ decays.
Figures \ref{Fig:mass}, \ref{Fig:acoplanar_a1rho} and \ref{Fig:acoplanar_a1a1} demonstrate the effect of each variation of the hadronic current on the mass and the acoplanarity angles of $a_1-\rho$ and $a_1-a_1$ decays respectively. \\

\begin{figure}[htb]
    \begin{center}
    {
        \includegraphics[width=6.8cm]{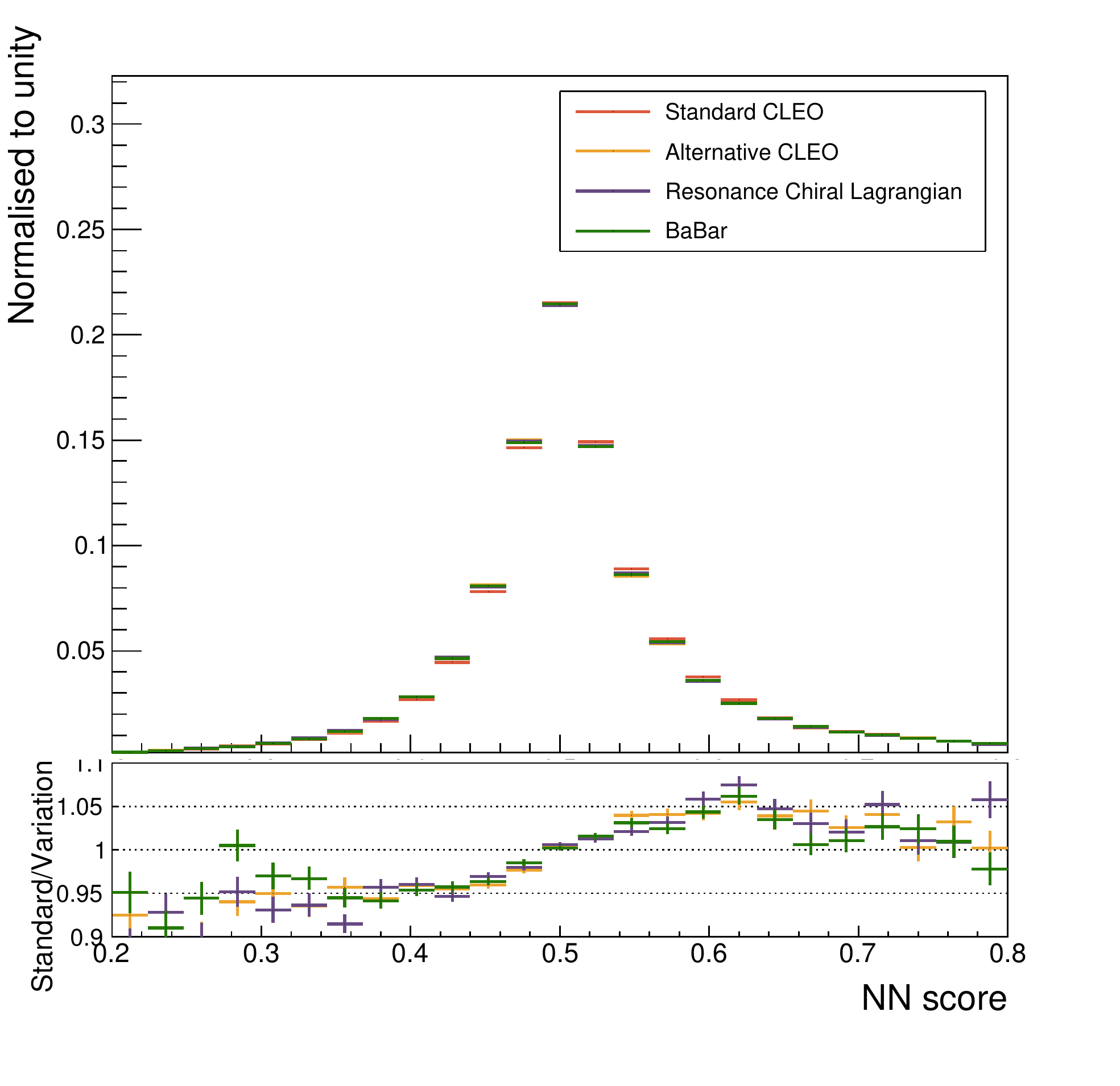}
        \includegraphics[width=6.8cm]{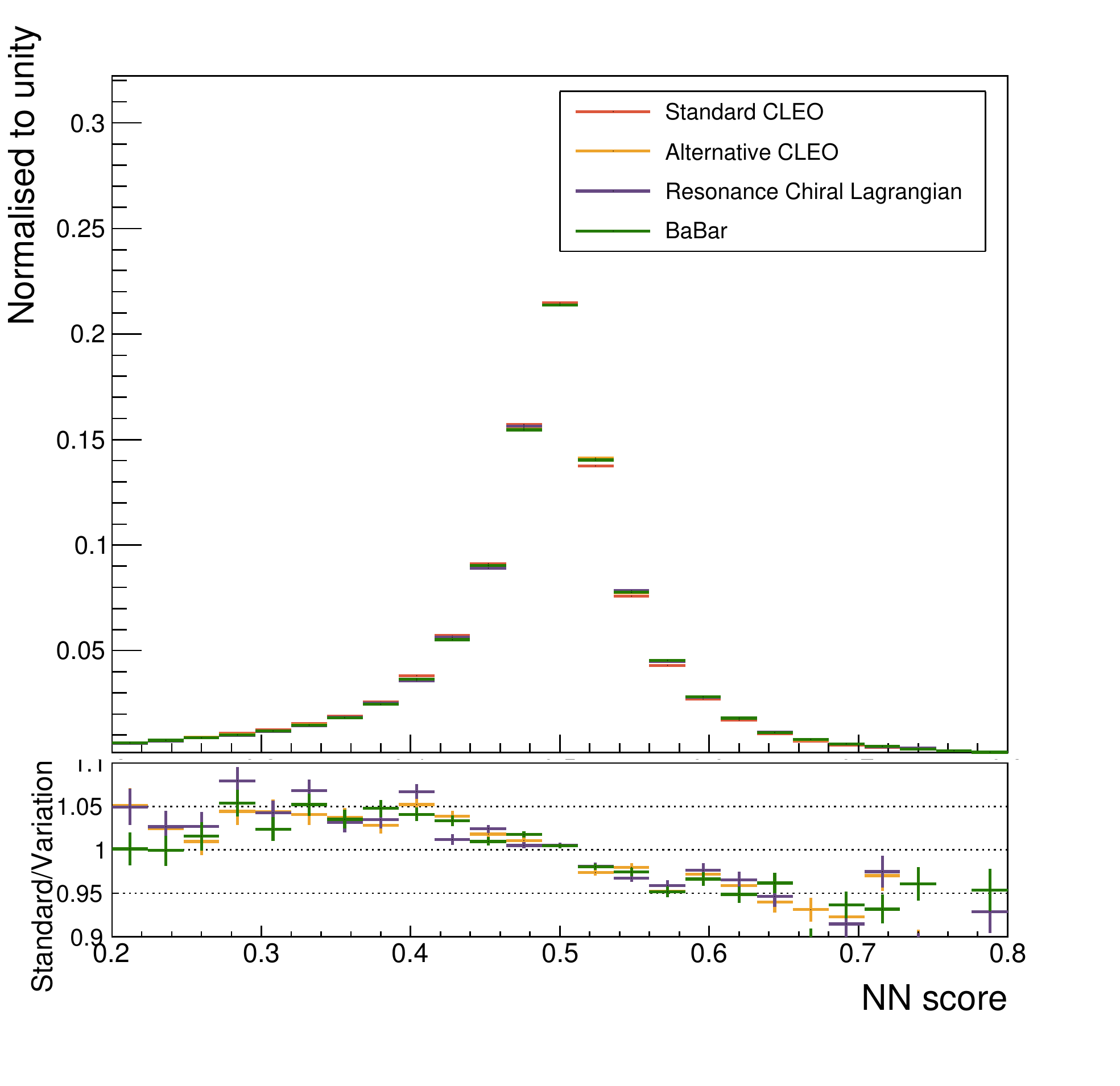}
    }
    \end{center}
    \caption{Distribution (for all four hadronic current samples) of the NN classifier. The lower plots show 
the ratios between the prediction for the Standard CLEO current (STD) and the other variations (RXL, ALT and BBR).
The left plot is for the scalar hypothesis and the right one, for the pseudoscalar.\label{Fig:NN}}
\end{figure}

Neural networks were trained on the default CLEO and then applied to samples simulated with variations of the hadronic currents mentioned above (with an equivalent sized sample). The AUC scores are presented for the $a_1-\rho$ and $a_1-a_1$ decay modes in Table \ref{table:currents}.\\ \\
\begin{table}
    \centering
    \begin{tabular}{|c|c|c|c|c|c|c|c|}
    \hline
    \multicolumn{4}{|c|}{Features}  & \multicolumn{1}{c|}{\multirow{2}{25 mm}{STD}} & \multicolumn{1}{c|}{\multirow{2}{25 mm}{R$\chi$L}} & \multicolumn{1}{c|}{\multirow{2}{25 mm}{ALT}} & \multirow{2}{25 mm}{BBR} \\ \cline{1-4}
    $\phi^{*}$ & 4-vec & $y_i$ & $m_i$ & \multicolumn{1}{c|}{} & \multicolumn{1}{c|}{} & \multicolumn{1}{c|}{} & \\
    \hline
    \multicolumn{8}{|c|}{$a_1-\rho$ Decays}\\
    \hline
        \cmark & \cmark & \cmark & \cmark & 0.604 & 0.604 & 0.603 & 0.603 \\
        \cmark & \cmark & \cmark & -      & 0.597 & 0.596 & 0.596 & 0.597 \\
        \cmark & \cmark & -      & \cmark & 0.604 & 0.604 & 0.604 & 0.604 \\
        -      & \cmark & -      & -      & 0.597 & 0.596 & 0.596 & 0.595 \\
        \cmark & \cmark & -      & -      & 0.597 & 0.596 & 0.596 & 0.595 \\
        \cmark & -      & \cmark & \cmark & 0.593 & 0.593 & 0.593 & 0.593 \\
        \cmark & -      & \cmark & -      & 0.582 & 0.579 & 0.580 & 0.578 \\
    \hline
    \multicolumn{8}{|c|}{$a_1-a_1$ Decays}\\
    \hline
        \cmark & \cmark & \cmark & \cmark & 0.567 & 0.563 & 0.564 & 0.564 \\
        \cmark & \cmark & \cmark & -      & 0.560 & 0.555 & 0.557 & 0.556 \\
        \cmark & \cmark & -      & \cmark & 0.568 & 0.564 & 0.566 & 0.566 \\
        -      & \cmark & -      & -      & 0.562 & 0.557 & 0.559 & 0.559 \\
        \cmark & \cmark & -      & -      & 0.562 & 0.557 & 0.559 & 0.559 \\
        \cmark & -      & \cmark & \cmark & 0.547 & 0.546 & 0.547 & 0.545 \\
        \cmark & -      & \cmark & -      & 0.537 & 0.534 & 0.535 & 0.533 \\
    \hline
    \end{tabular}
    \caption{Area under ROC curve. NN trained with $a_1-a_1$ decays of $\tau\tau$ system with standard CLEO current on exact MC
    sample. These NN are then tested on MC generated sample with alternative parameterisations of the hadronic currents.}
    \label{table:currents}
\end{table}
Little variation is observed between the different configurations of hadronic currents (of the order 1\% in the AUC score). 
This is encouraging for the expected stability of the discrimination power. 
With few exceptions, the variations in AUC score largely fall within 2 or 3$\sigma$ of the statistical uncertainty. 
The degradation due to training on samples with realistic experimental effects modelled is expected to impact the sensitivity more than the modelling of the 3 pion decays. This is indeed the case comparing Table \ref{table:baseline} and \ref{table:currents}.\\ \\
To illustrate the impact further, NN predictions can be assessed for each of the variations of hadronic currents (see Fig \ref{Fig:NN}). 
Clearly, there are only small deviations from the prediction based on the standard CLEO current. 
The trend in the lower plot is expected as the NN was trained on the standard CLEO current, the separation should be better.

\section{Potential Improvements}
The CP sensitivity of the $H\to\tau\tau$ decays has only been evaluated using the visible components of the $\tau$ lepton decays. 
As neutrinos escape experimental detection, the reconstruction of relevant kinematic variables, which can enhance discrimination power of the NN method, becomes challenging.
Theoretically though, this component is an important aspect to consider. 
The neutrino momentum enters the polarimetric vector and so partially defines the transverse polarisation of the tau decay products sensitive to the CP state of the Higgs boson. 
If one neglects the use of neutrino information completely in the inputs of the NN, there is a clear reduction in sensitivity with respect to the ``oracle" predictions (the limit of the approach) of 0.782 \cite{Jozefowicz:2016kvz}.\\ \\
The question of how to recover this lost information is a matter for future research, but will be briefly discussed here. 
As the $H\to\tau\tau$ system is largely boosted, the $\tau$ lepton decays fall into a regime in which the collinear approximation can be of some use. 
With the help of this approximation, one can largely constrain the longitudinal component of neutrino momentum (with respect to the $\tau$ momentum direction). 
Information from the decay vertex, impact parameters as well as the $\tau$ and $H$ boson masses may allow the remaining degrees of freedom to be sufficiently constrained such that the neutrino momenta can become useful.
\section{Conclusion}
Effects of detector resolution and $\tau$ decay modelling have been discussed with regard to the deep learning approach to the measurement of the Higgs boson CP state.
The impact of experimental effects were assessed and demonstrates, within a simplified Gaussian smearing of outgoing pion 4-momenta, the sensitivity of the NN remain largely stable.
The variations in the hadronic currents used to model $\tau$ decays show little impact on the separation of the NN. 
Therefore, the impact of $\tau$ decay modelling in $\tau\to a_1\nu$ can be considered to be minimal as a systematic effect on the deep learning approach to the measurement of the Higgs boson CP state.
\section*{Acknowledgements}
The author(s) would like to thank the support from funding agencies. Brian Le was supported by Australian Government Research Training Program Scholarship. Brian Le and Zbigniew Was were supported by European Union under the Grant Agreement PITNGA2012316704 (HiggsTools). Elisabetta Barberio and Daniele Zanzi are supported by the Australian Research Council through the Centre of Excellence for Particle Physics at the Terascale. Zbigniew Was and Elzbieta Richter-Was were supported by Polish National Science Centre under decisions UMO-2014/15/B/ST2/00049.\\ \\

\end{document}